\documentclass[a4paper,11pt]{article}
\usepackage{pos}
\usepackage{enumitem}

\usepackage{amsmath,amstext,amsthm,amsfonts,nicefrac}
\usepackage{cleveref}

\usepackage{tabularx, makecell}
\usepackage{booktabs} 
\usepackage{xspace}

\usepackage{hyperref} 
\usepackage{ulem}
\usepackage{multirow}

\newcommand\parbar[1]{\overset{\textbf{\fontsize{2pt}{2pt}\selectfont(---)}}{#1}}
\newcommand{\nue}                {$\nu_{e}$\xspace}

\newcommand{\numubar}        {$\bar{\nu}_{\mu}$\xspace}
\newcommand{\numu}             {$\nu_{\mu}$\xspace}

\let\OLDthebibliography\thebibliography
\renewcommand\thebibliography[1]{
  \OLDthebibliography{#1}
  \setlength{\parskip}{0pt}
  \setlength{\itemsep}{0pt plus 0.3ex}
}

\title{Uncertainties in modelling neutrino interactions for oscillation experiments }

\author*[a]{Stephen Dolan}

\affiliation[a]{CERN European Organization for Nuclear Research, CH-1211 Genéve 23, Switzerland}

\emailAdd{stephen.joseph.dolan@cern.ch}

\abstract{Accelerator-based neutrino oscillation experiments have the potential to revolutionise our understanding of fundamental physics, offering an opportunity to characterise charge-parity violation in the lepton section, to determine the neutrino mass ordering and to explore the possibility of physics beyond three-flavour neutrino mixing. However, as more data is collected the current and next-generation of experiments will require increasingly precise control over the systematic uncertainties within their analyses. It is well known that some of the most challenging uncertainties to overcome stem from our uncertain modelling of neutrino-nucleus interactions, arising because measured event rates depend on the neutrino interaction cross section in addition to any oscillation probability. The sources of these uncertainties are often related to subtle details of the pertinent nuclear physics, such as those of the target nucleus ground state, which are extremely difficult to control with sufficient precision. Confronting such uncertainties requires both state-of-art theoretical modelling and precise measurements of neutrino interaction event rates at experiment's near detectors, before oscillations are likely occur. These proceedings will briefly review the role of neutrino interaction systematic uncertainties in current and future measurements of neutrino oscillations. }

\FullConference{%
  Neutrino Oscillation Workshop 2022 \\
  4-11 September 2022\\
  Rosa Marina, Ostuni, Italy
}

\begin{document}
\maketitle

\section*{Neutrino interactions for neutrino oscillations}

A robust modelling of neutrino-nucleus interactions and a comprehensive set of associated systematic uncertainties are likely to be essential for the next generation of accelerator-based neutrino oscillation experiments to succeed. These experiments operate by producing a beam of GeV-scale predominantly \numu or \numubar, which impinges on both a “near” and “far” detector, usually positioned a few hundred metres and many hundreds or thousands of kilometres from the neutrino beam production point respectively. By inferring the degree of electron neutrino appearance and muon neutrino disappearance at the far detector, alongside their evolution as a function of neutrino energy, experiments can characterise neutrino oscillations. This can take the form of fits to parameters of the 3-flavour PMNS neutrino mixing framework or searches for physics beyond it. The crux of all such analyses is to infer neutrino oscillation probabilities as a function of neutrino energy from measured event rates. In the background-less case these can be written as: 
\begin{equation}
    N(E_{\nu}^{Rec}) = P_{\alpha \rightarrow \beta}(E_{\nu}^{True}) \Phi(E_{\nu}^{True}) \sigma(E_{\nu}^{True}) \epsilon(E_{\nu}^{True}) S(E_{\nu}^{True},E_{\nu}^{Rec}),
    \label{eq}
\end{equation} 
where $E_{\nu}^{True}$ and $E_{\nu}^{Rec}$ are the true and reconstructed incoming neutrino energy respectively; $P_{\alpha \rightarrow \beta}$ is the neutrino oscillation probability between flavours $\alpha$ and $\beta$; $\Phi$ is the unoscillated neutrino flux; $\sigma$ is the neutrino interaction cross section; $\epsilon$ is the detection efficiency; and $S$ the smearing matrix between true and reconstructed neutrino energy. It is clear that an inference of the neutrino oscillation probability can only be as accurate as the knowledge of each other term within the equation. For currently operating long-baseline experiments, T2K~\cite{t2k:neutrino} and NOvA~\cite{NOvA:2018gge}, the dominant systematic uncertainty on event rates are driven by uncertainties related to neutrino-nucleus interaction modelling, either via the $\sigma$ term or the way in which neutrino interaction model uncertainties can change $S$. For example, in recent analyses the predicted electron neutrino appearance event rate is subject to a 3.8\%/7.7\% uncertainty due to cross section related uncertainties out of a total 5.2\%/9.2\% for T2K/NOvA respectively~\cite{t2k:neutrino,NOvA:2018gge}. Whilst this level of uncertainty is acceptable for T2K and NOvA, which have $\sim$100 events in electron neutrino appearance and a few hundred events in muon neutrino disappearance far detector samples, it is not for the upcoming DUNE~\cite{DUNE:2020ypp} and Hyper-K~\cite{Hyper-Kamiokande:2018ofw} experiments, which expect a few thousand electron neutrino appearance events and a more than ten thousand in muon neutrino disappearance samples.

Experiments are already able to dramatically reduce uncertainties using measurements made at their near detectors, which measure event rates before oscillations are expected to occur. However, since near and far detectors cannot practically be identical, their differing acceptance, and in some cases differing nuclear targets, means that a neutrino cross-section model is needed to extrapolate constraints from the near detector to the far detector. More subtly, a perfect cancellation of systematic uncertainties in the near-to-far detector charged-current event rate ratios is prevented because the neutrino flavour composition is different due to neutrino oscillations. Since neutrino interaction cross sections evolve rapidly in energy at the GeV-scale, the charged-current flux-averaged cross sections at the near and far detector are significantly different. Moreover, measurements of \nue appearance at the far detector require modelling of the \nue interaction cross section, which is challenging to constrain at a near detector. Overall, whilst many components of neutrino interactions must be well understood, three particularly crucial aspects of their modelling for GeV-scale neutrino oscillation analyses can be identified, which are discussed in the following sections.

\section*{Energy dependence of neutrino-nucleus cross sections}

\begin{figure*}[htbp]
  \centering
\includegraphics[width=0.98\linewidth,trim={0 0mm 0 0.8mm},clip]{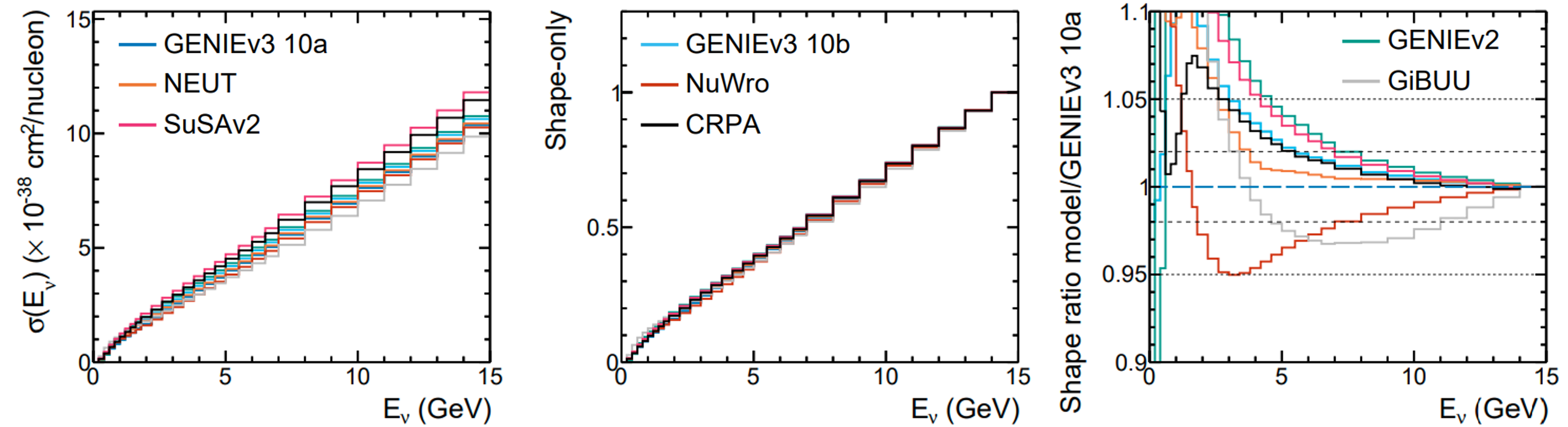}
  \caption{Comparison of the total \numu--Argon charged-current cross section predicted by a variety of event generators: as a function of neutrino energy (left); as a comparison of the shape of the same cross section, normalised such that the 14--15 GeV bin is 1 (middle); and as a comparison of the ratio of the shape-only model predictions with respect to GENIEv3 10a as a reference model (right). This figure and more details can be found in~\cite{Wilkinson:2022dyx}.}
  \label{fig:edep}
  \vspace{-5mm}
\end{figure*}

An accurate modelling of the energy dependence of neutrino-nucleus cross sections is needed to extrapolate interaction model constraints from our near to far detectors. A recent analysis~\cite{Wilkinson:2022dyx} explored how a variety of interaction models, including state-of-the-art theory and those most used by experimental collaborations, differ in their description of energy dependence. Fig.~\ref{fig:edep} shows how models differ in their prediction of the ratio of cross sections for different neutrino energies around the DUNE and Hyper-K flux peaks ($\sim$2 GeV and 0.6 GeV respectively) by more than the statistical uncertainty expected in far detector samples given the aforementioned expected event rates. The energy dependence of interaction cross sections may be better constrained using future near detectors capable of taking data at different positions with a neutrino beam, thereby seeing predictably different neutrino energy spectra~\cite{nuPRISM:2014mzw}.

\section*{Neutrino energy reconstruction}

\begin{figure*}[htbp]
  \centering
\vspace{-4mm}
\includegraphics[width=0.95\linewidth,trim={0 0.5mm 0 1mm},clip]{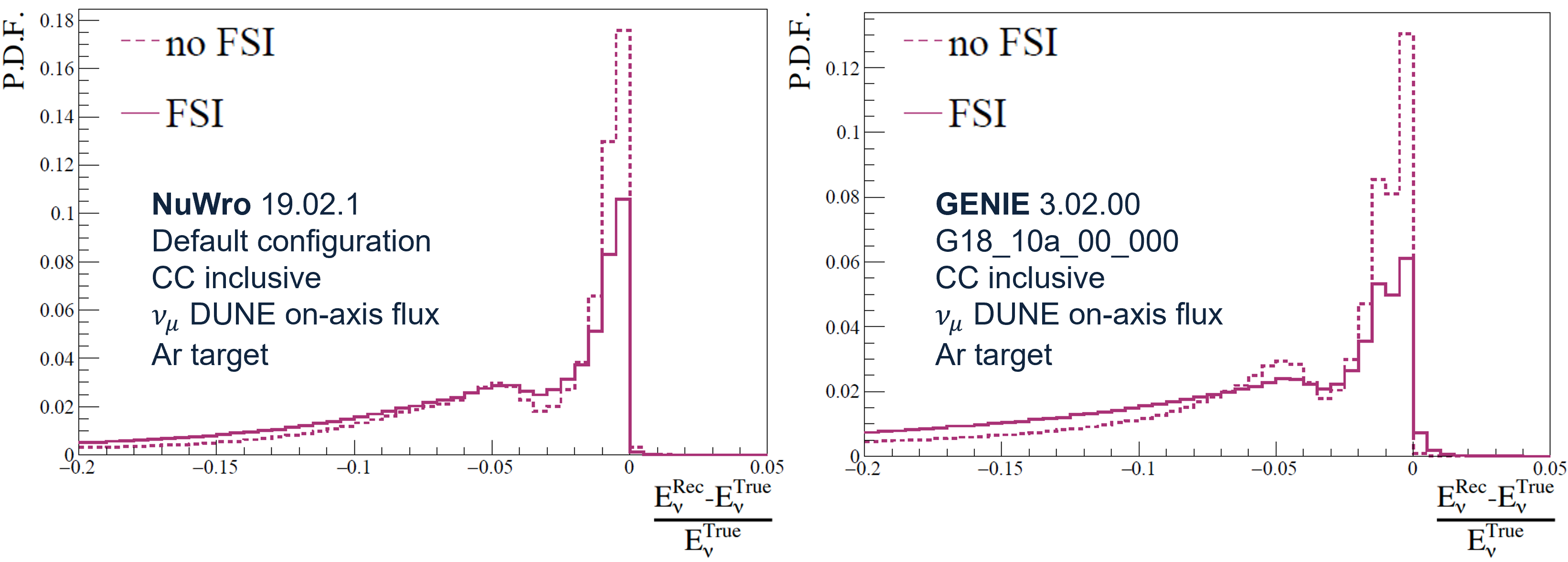}
  \caption{A comparison of GENIE and NuWro, two widely-used neutrino interaction event generators, estimations of neutrino energy reconstruction bias in a DUNE-like configuration with and without the modelling of FSI. It is clear that the impact of FSI on neutrino energy reconstruction bias is different for the two models. This figure and more details can be found in~\cite{Valentino:2825404}.}
  \vspace{-3mm}
  \label{fig:fsi}
\end{figure*}

It is crucial to accurately model the mapping between true and reconstructed neutrino energy (i.e. $S$ in Eqn.~\ref{eq}) to allow an accurate inference oscillation probability shape from measured event rates. Different experiments reconstruct neutrino energies using different techniques, but for NOvA and DUNE one particular challenge in modelling this mapping stands out. NOvA and DUNE reconstruct neutrino energy by calorimetrically summing energy deposits observed in their detectors. They will typically not see a significant fraction of the energy deposited by neutrons inside the detectors, and so a model is needed to estimate the fraction of unseen energy these carry away. Modelling this is extremely challenging. Among many other processes, neutrons can be produced by the re-interaction of hadrons inside the nucleus as they undergo ``Final state interactions'' (FSI). FSI are usually modelled using semi-classical methods which can give significantly different predictions depending on the approximations made. An example of this is shown in Fig.~\ref{fig:fsi}, which demonstrates that differing approaches to modelling FSI can lead to significantly different predictions of how FSI changes $S$. The use of more advanced FSI models such as those discussed in~\cite{Ershova:2022jah,Franco-Patino:2022tvv} may reduce the scope for large FSI-related uncertainties in future neutrino oscillation analyses. 

\section*{Electron-muon neutrino cross-section differences}

Differences in the $\parbar{\nu}_e$ and $\parbar{\nu}_{\mu}$ cross sections are crucial to allow a near detector to constrain the cross section for \nue appearance measurements. Assuming lepton universality, the only difference in the cross sections is from the differing lepton mass. However, the way in which lepton mass change the cross section depends on a variety of poorly understood physics processes. Recent work has shown that alterations to the nuclear model~\cite{Tomalak:2021hec} and the consideration of radiative corrections to cross section calculations~\cite{Dieminger:2023oin} can alter the $\parbar{\nu}_e$ to $\parbar{\nu}_{\mu}$ cross section ratio by $\sim$2-4\%, which is not negligible compared to expected statistical uncertainties for Hyper-K and DUNE. 
\vspace{-1mm}
\section*{Outlook}

It is clear that uncertainties on the neutrino interactions physics most pertinent to future oscillation experiments are far from under control and that, as more data is gathered, experiments must consider an increasingly wider range of analysis failure modes due to a mismodelling of neutrino interactions. Efforts to do this are underway, but there remains significant experimental and theoretical challenges toward developing a neutrino interaction model suitable for the late-stage analyses of the next generation of neutrino oscillation experiments. However, it is encouraging to note the enormous progress in neutrino interaction modelling and measurements within the last decade as well as the wide variety of ongoing efforts.

\vspace{-11mm}

\small

\bibliographystyle{apsrev}
\bibliography{biblo}


\end{document}